\documentclass[Conference]{IEEEtran}
\IEEEoverridecommandlockouts
\usepackage{cite}
\DeclareUnicodeCharacter{2121}{Tel}
\usepackage{amsmath,amssymb,amsfonts}
\usepackage{algorithmic}
\usepackage{url}
\usepackage{graphicx}
\usepackage{textcomp}
\usepackage{xcolor}
\usepackage{caption}
\usepackage{booktabs} 
\usepackage{float}
\def\BibTeX{{\rm B\kern-.05em{\sc i\kern-.025em b}\kern-.08em
    T\kern-.1667em\lower.7ex\hbox{E}\kern-.125emX}}

\begin{document}
\title{
Economic Impact Assessment of Denial-of-Service and Time-Delay Attacks on Advanced Metering Infrastructure \\
} 

\vspace{5mm}
\author{
\begin{minipage}[t]{0.32\textwidth}
\centering
\textbf{Lais Oliveira Krohl}\\
\textit{Program of Electrical and Computer Engineering}\\
\textit{Wentworth Institute of Technology}\\
Boston, MA, USA\\
deoliveiral2@wit.edu
\end{minipage}
\hfill
\begin{minipage}[t]{0.32\textwidth}
\centering
\textbf{Saurav Basnet}\\
\textit{Program of Electrical and Computer Engineering}\\
\textit{Wentworth Institute of Technology}\\
Boston, MA, USA\\
basnets@wit.edu
\end{minipage}
\hfill
\begin{minipage}[t]{0.32\textwidth}
\centering
\textbf{Ioannis Zografopoulos}\\
\textit{Engineering Department}\\
\textit{University of Massachusetts Boston}\\
Boston, MA, USA\\
i.zografopoulos@umb.edu
\end{minipage}
\vspace{-0.8cm}
}

\maketitle

\begin{abstract}
Advanced Metering Infrastructure (AMI) plays an important role in modern power systems by providing near real-time consumption data for operational planning and load estimation. Despite this, AMI communication networks are vulnerable to certain \textcolor{black}{cyberattack} types, such as denial-of-service (DoS) and time delay attacks (TDAs), which may compromise data availability. This paper evaluates the impact of these attack types on AMI communication using a Mininet-based simulation environment that \textcolor{black}{implements a communication model inspired by the Distributed Network Protocol 3 (DNP3)} for communication among smart meters, data concentrators, and utility systems. Different cyberattack scenarios are analyzed, including communication delay, packet loss, combined degradation rate ($D_{rate}$), and concentrator failure, to evaluate their effects on network performance and data availability. The results indicate increased latency, higher communication unavailability rates, and reduced data reliability under attack conditions. In addition, the study correlates how  communication degradation can result in potential uncertainties in energy demand estimation and provides estimates of the associated financial impacts on energy procurement, using electricity pricing data from the Independent System Operator--New England (ISO--NE). The experimental results reveal that a 500 ms delay increased the average \textcolor{black}{RTT (Round-trip time)} from \textcolor{black}{0.22 ms} to 69.5 ms, while a 50\% packet loss scenario resulted in 87 timeout events and a 5.31\% communication degradation rate. Under a regional-scale deployment of 10,000 smart meters, this degradation \textcolor{black}{could cause an} estimated economic impact of \$1,009/day under normal market conditions and up to \$5,097/day during high-price events.
\end{abstract}


\vspace{-0.2cm}
\section{Introduction} \label{s:Intro}

The growing digitalization and interconnectivity of modern power systems has introduced new cybersecurity vulnerabilities in critical energy infrastructure~\cite{zografopoulos2025cyber}. Incidents around the world have demonstrated the severity of these threats. For instance, in December 2015, a cyberattack using the BlackEnergy malware caused power outages for approximately $225,000$ customers across three energy distribution companies in Ukraine, marking the first confirmed large-scale blackout caused by a cyberattack \cite{noauthor_cyber-attack_2021}. Beyond physical disruptions, cyberattacks targeting the communication and energy metering infrastructure can also generate significant economic consequences\cite{zografopoulos2023distributed, zografopoulos2025event, topallaj2025impact}. Namely, electricity theft and data manipulation in smart grid networks are estimated to cost the industry up to $\$96$ Billion annually worldwide~\cite{jakaria_safety_2019}.

AMI plays an essential role in modern power systems by providing near real-time consumption data~\cite{petrusevski_novel_2014, muthamizh_selvam_initiatives_2016}. The AMI information is used by utilities to estimate energy demand and guide operational decisions related to the following days' and weeks' energy supply. \textcolor{black}{In contrast}, when AMI data arrives with significant delays, e.g., due to TDAs ~\cite{yi_denial_2014, zhe_dos_2020}, or suffers from DoS attacks, the accuracy of load forecasts can be significantly affected. Consequently, utilities face difficulties in energy planning and may have to purchase surplus energy in the real-time market, often introducing significant costs.

Given the potential impacts of the aforementioned attacks and the weak cybersecurity standpoint of AMI devices, many recent research studies have focused on smart grid security. In general, existing studies can be grouped into two categories. The first category includes works focused on general AMI security threats, i.e., device-level, and their possible countermeasures \cite{khoei_comprehensive_nodate,achaal_study_2024}, with some proposing detection mechanisms such as collaborative intrusion detection systems targeting false data injection attacks \cite{liu_collaborative_2015}. On the other hand, the second category focuses specifically on availability attacks, i.e., communication network and protocol levels, including DoS and TDAs targeting AMI components and smart grid networks \cite{de2026let, zhe_dos_2020, zhang_modeling_2021, de2026quic, rath2022behind, lou_assessing_2020}.

Although these studies demonstrate the severity of such attacks, showing that even modest delays can degrade meter data delivery and compromise AMI meter streams used for load forecasting and day-ahead planning, little attention has been given to how these communication degradations can translate into economic consequences. Understanding the financial impacts caused by delays of consumption data used for demand forecasting remains an operational research question.

This paper investigates the impacts of communication degradation associated with DoS and TDAs, as well as operational failures, targeting AMI networks through simulations carried out using the Mininet network emulation platform that enables the virtual modeling of an AMI infrastructure. The study focuses on analyzing how these communication degradations can affect the accuracy of demand forecasting and, consequently, introduce additional costs associated with energy procurement in the real-time market.

The main contributions of this work are as follows:
\begin{itemize}
    \item Development of a Mininet-based AMI communication architecture that \textcolor{black}{employs a DNP3-inspired communication model over UDP} to relay meter data. 
    \item Implementation of cyberattack-inspired communication degradation scenarios including: \emph{i)} delay, \emph{ii)} packet loss, \emph{iii)} combined degradation, and \emph{iv)} data concentrator failure, to evaluate their effects on AMI availability.
    \item Quantitative estimation of the economic consequences associated with AMI communication degradation, derived from experimentally observed impairment rates and real-time electricity pricing data from the Independent System Operator ISO--NE.
\end{itemize}

The remainder of this paper is organized as follows. Section II provides background on AMI, DNP3, and day-ahead electricity market. Section III describes the proposed methodology, covering the AMI system modeling, \textcolor{black}{attack} scenarios, performance metrics, and the economic evaluation framework. Section IV reports the experimental results and \textcolor{black}{discusses} their implications for communication reliability and the associated economic costs. Finally, Section V concludes the paper and outlines directions for future research.

\section{Background and Preliminaries} \label{s:Background}

The following three subsections provide the preliminary definitions and contextual information essential for the security analysis presented in this work. 

\vspace{-0.5cm}
\subsection{Advanced Metering Infrastructure}
According to IBM, AMI is \emph{``an integrated, fixed-network system that enables two-way communication between utilities and customers"}~\cite{IBM_AMI}. The architecture is typically organized into four main components. \textcolor{black}{Smart meters are installed at customer premises and record energy-consumption data at intervals commonly ranging from $15$ minutes to one hour \cite{kuzlu_communication_2014}. Depending on the utility configuration, these readings may be retrieved several times per day through the AMI infrastructure. Data concentrators collect and aggregate measurements from multiple meters before forwarding them to utility back-end systems}. 
Base station switches handle the communication between the data concentrators and the utility infrastructure. Finally, the utility servers receive, validate, and processes all incoming data used for operational decision-making~\cite{muthamizh_selvam_initiatives_2016}. 

\vspace{-0.5cm}
\subsection{Distributed Network Protocol 3}
DNP3 is the most widely used communication standards in industrial control systems, particularly in SCADA (Supervisory Control and Data acquisition) environments and smart grids~\cite{zografopoulos2020derauth}. The DNP3 protocol was developed to provide reliability, with features that ensured data integrity and measurements. Nevertheless, cybersecurity was not a primary consideration when the protocol was created and implemented. To address these security limitations, newer versions of DNP3, such as DNP3-SAv6, have been introduced. Nonetheless, already deployed and legacy devices with limited computational resources might not be able to support the secure DNP3 versions, exposing them to known vulnerabilities~\cite{zografopoulos2023distributed}. Such vulnerabilities include message interception, replay attacks, TDAs, false data injection attacks, and DoS\cite{ospina2020trustworthy, zografopoulos2020special, achaal_study_2024}.

\subsection{Day-ahead Electricity Markets }
In deregulated electricity markets such as the ISO-NE, utilities are required to forecast their expected energy demand for the following day and submit bids in the day-ahead market, where energy is traded at pre-negotiated prices \cite{noauthor_day-ahead_nodate}. This process relies on the \textit{quality} and \textit{timeliness} of the consumption data collected through AMI systems~\cite{dewangan_load_2023}. When AMI measurements are delayed, lost due to cyberattacks, or maliciously manipulated, utilities must rely on outdated or inaccurate information. This can result in an underestimation of future energy demand. As a result, utilities may be forced to procure additional electricity in the real-time market, where prices are typically more volatile, less predictable, and typically higher than those in day-ahead markets. \cite{noauthor_day-ahead_nodate}. 
\section{Methodology} \label{s:Method}

\subsection{AMI Network Architecture}

\subsubsection{AMI System Modeling}
All experiments were executed on a Linux virtual machine running Ubuntu 20.04.1 LTS under Oracle VirtualBox, configured with 1 virtual CPU and 1 GB of allocated memory. Mininet, a network emulator, was used as the network emulation platform due to its ability to reproduce configurable AMI communication topologies with low computational overhead \textcolor{black}{\cite{lantz_network_2010}.} Since Mininet configurations do not persist after a session is terminated, the entire topology was defined using Python scripts specifying all hosts, switches, links, and their respective interfaces. This approach guarantees that the same environments are instantiated in every test run, maintaining a consistent experimental setup. 

\begin{figure}[!t]
    \centering
    \captionsetup{font=normalsize}
    \includegraphics[width=0.65\columnwidth]{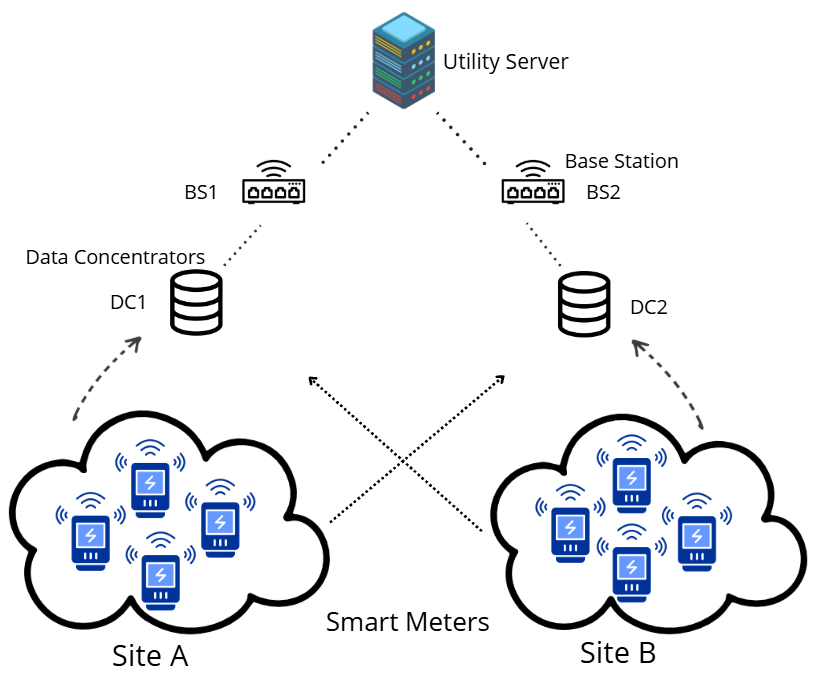}
    \caption{Proposed dual-redundant communication architecture.}
    \label{fig:Arch}
    \vspace{-0.5cm}
\end{figure}

The topology, as seen in Fig. \ref{fig:Arch}, comprises eight smart meters, two data concentrators (DC1 and DC2), two base stations (BS1 and BS2), and one utility server. Smart meters, data concentrators, and the utility server were modeled as Mininet hosts with static IP addresses, 
while base stations were configured as switches responsible for packet forwarding. Since the architecture is organized into two sites interconnected through multiple redundant paths, all available communication paths between devices were explicitly declared in the scripts. These exact path declarations 
ensure that the same path configurations are reproduced in every experiment.

Redundant communication paths were incorporated to enable resilience testing under device failure conditions and to replicate realistic utility networks. During normal operation, both data concentrators poll all smart meters.
Smart meters in Site A (M1–M4) are assigned to DC1 as their primary concentrator with DC2 as backup, while Site B meters (M5–M8) follow the inverse arrangement. This design ensures communication continuity when one concentrator becomes unavailable and serves as the basis for the \textit{DC-Down} scenario described in Section \ref{s:attackModel}.

\subsubsection{Communication Model}
A communication model inspired by the DNP3 protocol was implemented to reproduce the ``\textit{master~--~outstation}" behavior characteristic of AMI systems. Smart meters were configured as outstations, data concentrators as masters, and the utility server as the central monitoring node. Communication was implemented over User Datagram Protocol (UDP) sockets to preserve lightweight request–response behavior and provide direct control over timeout handling and retransmission logic.

Since both concentrators poll every smart meter, duplicate measurements regularly arrive at the utility server. To avoid duplicates, a filtering mechanism using timestamps and measurement content comparison is implemented at the utility server. In this mechanism, when two readings sharing the same timestamp and content arrive, one of the gets discarded.

\textcolor{black}{Although DNP3 can operate over different communication media, IP-based implementations typically use TCP, which provides in-order data delivery and retransmission of lost segments 
\cite{eddy_transmission_2022}. DNP3 also supports data-link and application-layer confirmations, depending on the system configuration \cite{clarke_cp_eng_practical_2004}. In this work, however, a DNP3-inspired communication model was implemented over UDP. The model does not retransmit failed requests, and 
if a response is not received within the 1-s timeout, the request is recorded as a timeout, and a new poll occurs only in the following cycle. Therefore, the results reflect the behavior of this specific model and should not be directly generalized to DNP3/TCP systems. In such systems, retransmissions may recover part of the packet loss, although they may increase latency and network traffic.}

\subsubsection{Polling Parameters and Data validation}
Timing parameters were established based on the behavior observed during baseline operation, as shown in Fig. 2. A timeout threshold of 1 second was defined as the maximum waiting time between the issuance of a $Read\_Meter$ request and the arrival of the corresponding response. Requests exceeding this limit are registered as timeout events. A polling interval of 3 seconds was set between consecutive polling cycles to prevent request overlap while maintaining regular data collection. During each polling cycle, the data concentrator sequentially issues $Read\_Meter$ requests to all smart meters. The duration of each cycle depends on the cumulative response time of all devices. Therefore, the 3  second interval represents the minimum waiting period between cycles rather than a fixed total cycle duration. If the polling execution exceeds this interval due to communication delays or timeout events, the next cycle starts immediately after the current cycle ends. As a result, the total number of recorded polling transactions may vary across scenarios depending on network conditions.

\begin{figure}[!t]
    \centering
    \includegraphics[width=\columnwidth]{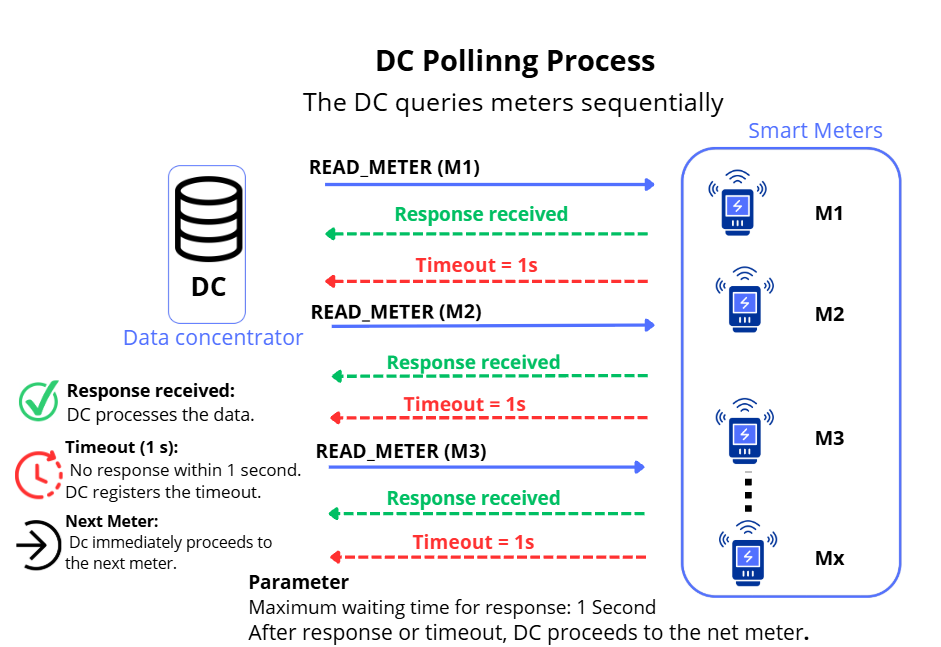}
    \caption{DNP3-inspired master-outstation polling mechanism 
    between the data concentrator and smart meters.}
    \label{fig:poll}
    \vspace{-0.5cm}
\end{figure}

\textcolor{black}{The simulation used a 3 seconds interval as a time-compression parameter to generate repeated request-response exchanges during shorter emulation runs. This experimental setting does not represent typical AMI reporting intervals. 
Longer reporting intervals and utility-scale traffic may produce different congestion and temporal loss patterns. Future work will evaluate these conditions. }  

No retransmission attempts were performed within the same polling cycle. Failed requests were retried in subsequent cycles, reflecting practical AMI polling systems, where missed measurements are recovered in future  intervals. Each request and response event was associated with a timestamp, enabling the: \emph{i)} calculation of the $Round~Trip$ time ($RTT = t_{response} - t_{request}$), \emph{ii)}  detection of communication failures when exceeding timeout limits, and \emph{iii)} identification of duplicate measurements received by both data concentrators. \looseness=-1

\vspace{-0.5cm}
\subsection{Threat and Attack Models}

\subsubsection{Threat Model}
Five degradation scenarios are implemented using the Linux Traffic Control utility (tc netem) ~\cite{hemminger_network_nodate}, which allows controlled insertion of delay and packet loss on individual communication links of the emulated topology. The scenarios are designed to represent realistic AMI attack conditions across a progressive severity spectrum. Importantly, none of the scenarios is configured to force a total system collapse, and rather than causing an instantaneous interruption, the injected impairments produce gradual and measurable degradation levels suitable for performance characterization.

\subsubsection{Attack Models} \label{s:attackModel}
The following five attack models are investigated in this study.
\paragraph{Time-Delay Attack}A fixed delay of 500 ms is applied to selected communication links. This 500 ms delay,  represents a substantial fraction of the 1 second timeout threshold and introduces measurable RTT increases without  triggering systematic timeouts that can cause communication degradations.

\paragraph{Packet Loss Attack (20\%)}A $20\%$ packet loss rate is introduced, causing one in every five packets to be dropped. This scenario preserves overall system functionality but generates intermittent timeouts and data gaps, reducing the availability and reliability of the aggregated information.

\paragraph{Packet Loss Attack (50\%)}A $50\%$ packet loss rate is applied, causing one in every two packets to be lost. This scenario substantially increases timeout frequency, introduces data uncertainty, and reduces measurement availability, suggesting that operation is approaching the limit supported by the communication architecture.

\paragraph{Combined Communication Degradation}Delay and packet loss of 20\% were simultaneously applied to evaluate cumulative degradation under attack conditions, in which the combined effects increase timeout occurrence and reduce data availability while the system remains partially operational.

\paragraph{Data Concentrator failure (DC-down)}One data concentrator is intentionally disabled to emulate an operational failure of hardware, software, or network link. 
This scenario quantifies the value of the redundant communication architecture by verifying whether measurement visibility is preserved through the remaining concentrator when part of the infrastructure becomes unavailable.

\vspace{-0.5cm}
\subsection{Performance Metrics for Communication Degradation}
The following performance metrics were collected for each scenario: \textit{RTT, timeout count, successful measurement count, duplicate count,} and \textit{measurement availability}. Since timeouts are related to information availability, the number of timeouts registered within a given number of polling requests is used to characterize communication reliability. The communication degradation rate is defined as, $D_{rate} = \frac{N_{timeout}}{N_{requests}} $,
where $N_{timeout}$ represents the number of failed communication attempts and $N_{requests}$ is the total number of polling transactions recorded during the test period. Baseline tests were executed for 30 seconds, while attack scenarios were executed for 60 seconds. These durations were selected to provide sufficient communication activity for statistical observations, since the number of polling transactions may vary depending on communication delays and timeout events.
\vspace{-0.2cm}
\subsection{Economic Impact Estimation }
The measured $D_{rate}$ is extrapolated to a regional AMI deployment of 10,000 smart meters. The economic impact of communication degradation was estimated using, $Impact = N \times E \times D_{rate} \times P$, 
where $N$ is the number of affected meters, $E$ is the average daily energy consumption per meter, $D_{rate}$ is the measured communication degradation rate, and $P$ is the ISO-NE real-time electricity price. This formulation provides a simplified estimate of the operational cost associated with degraded AMI communication reliability at scale.

\vspace{-0.2cm}

\section{Results and Discussion} \label{s:Results}

\subsection{Communication Performance Under Attack}
To evaluate the impact of communication degradation on AMI performance, attack scenarios are executed and compared against a reference scenario without communication impairment. Under normal operating conditions, the reference scenario presents stable communication behavior, with an average RTT of approximately 0.22 ms and only one timeout event is observed during execution. 

In the first time-delay attack scenario, where a 500 ms communication delay is introduced, communication latency and the average RTT increase to approximately 69.5 ms. Despite this increase in latency, only one timeout event is recorded, which indicates that AMI systems can tolerate moderate communication delays without immediate service interruption. \looseness=-1

In the second packet loss attack scenario, communication reliability is impacted. Under a packet loss rate of $20\%$, 28 timeout events are recorded, accompanied by an average RTT of approximately 47 ms. To further investigate packet loss-related conditions, a more severe scenario with 50\% packet loss is implemented, resulting in a higher timeout occurrence of 87 recorded events. These results demonstrate that packet loss attacks 
can lower data availability.

The combined degradation attack, consisting of simultaneous 500 ms communication delay and $20\%$ packet loss, produced the worst communication impairment among all scenarios. A total of 104 timeout events are recorded, with an average RTT of approximately 51.4 ms. Notably, the combined scenario produced a higher average RTT than the packet loss only scenarios, suggesting a compounding impact of simultaneous delay and packet loss on communication performance. \looseness=-1

\begin{figure}[t]
\centering
\includegraphics[width=0.85\columnwidth]{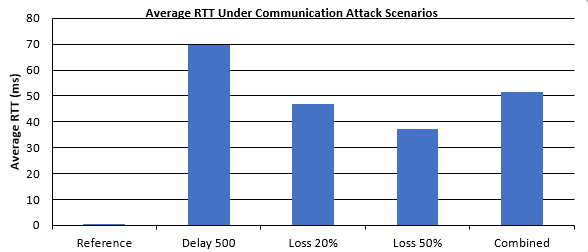}
\caption{Average Round-Trip Time (RTT) observed under different communication degradation scenarios.} 
\label{fig: RTT}
\vspace{-0.5cm}
\end{figure}

Fig. \ref{fig: RTT} presents the average RTT across all 5 scenarios evaluated in this study. The time-delay attack presented the highest average response time among the performed simulations, highlighting the direct impact of added latency on the AMI network. These results suggest that an increase in RTT does not guarantee an immediate communication failure, since only one timeout event is recorded in this scenario. Therefore, delay-based attacks primarily affect communication latency without immediately compromising data availability

\begin{figure}[t]
\centering
\includegraphics[width=0.85\columnwidth]{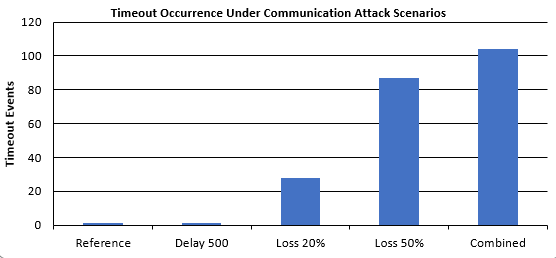}
\caption{Timeout occurrence under different scenarios.}
\label{fig:timeout}
\vspace{-0.5cm}
\end{figure}

As shown in Fig. \ref{fig:timeout}, the number of timeout events increases as the attack becomes more severe. The reference and delay scenarios produce only 1 timeout each, while packet loss raises this number to 28 at 20\% loss, 87 at 50\% loss, and 104 in the combined scenario. This shows that packet loss, and not delay, is the main cause of communication failures. The 500 ms delay increases the average RTT but stays below the 1 s timeout limit, so it rarely causes a timeout. It is also worth noting that most requests are not affected: the median RTT stays around 0.18 ms in every scenario, even when the average rises to 47 - 69 ms. This means the degradation is concentrated on a few affected links rather than spread across the whole network. The combined scenario shows both the highest number of timeouts (104) and the highest average RTT (51.4 ms), since delay and packet loss add up when applied together.\looseness=-1

\vspace{-0.3cm}
\subsection{Communication Resilience}
A data concentrator failure scenario is executed and the results indicate that the implemented redundant architecture remains operational even after one of the concentrators becomes unavailable. Since all smart meters continuously transmit their measurements to both concentrators, communication continuity is preserved through the remaining operational communication paths, preventing complete measurement loss. With both concentrators operational, only 1 timeout event is registered out of 445 \text{Read\_Meter} requests (a 0.22\% timeout rate). When one concentrator becomes unavailable, the number of timeout events increases to 104 out of 2{,}311 requests (4.50\%), the majority of which (102) occur on the affected concentrator. Although this represents an approximately twenty-fold increase in the timeout rate, the overall availability of measurements remains high. These observations highlight that communication redundancy mechanisms can significantly reduce the risk of data collection failures and maintain adequate levels of data availability for demand estimation processes~\cite{romero2026resilience, romero2026multidimensional}. 

\vspace{-0.2cm}
\subsection{Economic Impact Analysis of Communication Degradation}
Although the communication experiments are conducted using a small-scale AMI topology composed of only eight smart meters, the primary objective is to reproduce the functional architecture commonly observed in AMI communication networks deployed in electric power distribution systems. The reduced-scale topology is selected due to computational feasibility limitations and the complexity of reproducing high-fidelity large scale AMI environments.

To evaluate the potential operational and economic consequences under realistic conditions, the observed communication degradation behavior is extrapolated to a representative regional-scale scenario, as described in the methodology. A case involving 10,000 affected smart meters is considered to estimate the potential economic consequences of communication degradation caused by cyberattacks. This value is chosen to replicate the ongoing AMI modernization initiatives in the New England region. For instance, Eversource Energy received approval from the Massachusetts Department of Public Utilities to deploy more than 1.5 million smart meters across 159 communities in Massachusetts \cite{noauthor_grid_nodate}. The 10,000 meter scenario therefore represents a conservative regional scale estimate when compared to actual deployment sizes.

This analysis considers that delayed, unavailable, or degraded measurements may reduce demand forecasting accuracy, increasing discrepancies between forecasted and actual electricity demand. Under such conditions, utilities may be required to procure additional energy through the real-time electricity market, which is generally characterized by higher price volatility than the day-ahead market.

Based on the economic model presented in Section III, the \textit{Impact} of communication degradation is converted into an estimated daily operational cost. The parameters of the economic model are summarized in Table~\ref{tab:economic_model}. The cost estimate assumes that the degradation rate applies to the share of energy that the utility must purchase in the real-time market.

\begin{table}[t]
\small
\centering
\caption{Parameters of the Economic Model}
\label{tab:economic_model}

\resizebox{\columnwidth}{!}{
\begin{tabular}{|p{5cm}|c|c|}
\hline 
Parameter & Symbol & Value \\
\hline \hline
Affected meters & N & 10,000 \\
Average daily consumption per meter & E & 0.0320 MWh/day \\
Customer mix (residential/commercial) & -- & 90\% / 10\% \\
Communication Degradation Rate & $D_{rate}$ & 5.31\% (87/1638) \\
\hline
\end{tabular}
} \vspace{-0.2cm}
\end{table}

The value E = (0.0320 MWh/day) represents the average daily energy consumption adopted per affected meter. This value is derived from a weighted customer mix, considering (90\% residential, 10\% commercial ), consistent with Form EIA-861, Electric Power Sales, Revenue, and Average Price (2024) ~\cite{noauthor_electric_nodate}. Residential customers in Massachusetts average approximately 0.0190 MWh/day (EIA Table T5.A) and commercial customers approximately 0.1490 MWh/day (EIA T5.B), yielding $(0.9 \times 0.0190) + (0.10 \times 0.1490) = 0.0320$ MWh/day~\cite{noauthor_electric_nodate}.  $D_{rate} = 5.31\%$ corresponds to 87 timeouts events over 1,638 polling transactions recorded in the 50\% packet-loss scenario. In practice, utilities can mitigate part of the effect of missing data through interpolation, historical load profiles, or other data estimation methods, which are not addressed in this study. Therefore, the reported values should be interpreted as a conservative order-of-magnitude estimate rather than an exact prediction, since the actual impact depends on the operational methods adopted by each utility.

\indent Table~\ref{tab:economic_impact} presents the estimated daily impact for the two market conditions considered. Under reference conditions, a value of (P = \$59.38/MWh \cite{noauthor_december_2018})is adopted, based on ISO-NE data obtained from the \textit{Day-Ahead Energy Market Hourly LMP Report} \cite{noauthor_day-ahead_nodate}. Although different \textit{Network Nodes} present distinct values, this analysis uses the value associated with PNode LD.ALLINGS 13.8, located at Allings Crossing, West Haven, CT, with a voltage class of 13.8 kV and belonging to the Southwest Connecticut (SWCT) load zone. Applying this price to the scenario with 10,000 affected meters, together with the remaining parameters of the economic model, the estimated impact reaches approximately \$1,009 per day. This value represents a conservative lower bound, since it is derived exclusively from the experimentally observed results.

\begin{table}[t]
\centering
\normalsize
\caption{Estimated Economic Impact by Scenario}
\label{tab:economic_impact}

\resizebox{\columnwidth}{!}{
\begin{tabular}{|p{6.5cm}|p{2cm}|p{2cm}|}
\hline
Scenario & RT Price (P) & Est. Cost/day \\
\hline \hline
Baseline (ISO-NE) & \$ 59.38/MWh & $\approx$ \$ 1,009 \\
Worst-Case (Cold Event 2017-18) & \$ 300 /MWh & $\approx$ \$ 5,097\\
\hline
\end{tabular}
}
\vspace{-0.5cm}
\end{table}

To illustrate the potential severity of a coordinated cyberattack during periods of extreme electricity prices, a conservative worst-case price of \$300/MWh is also adopted. This value is based on the December 2017- January 2018 \textcolor{black}{cold-weather} event, when real-time Hub Locational Marginal Prices (LMPs) at the New England Hub exceeded \$300/MWh and reached approximately \$370/MWh \cite{noauthor_december_2018}. In this analysis, only the energy price is changed, while all other model parameters are kept constant. Under this assumption, the estimated impact increases to approximately \$5,097 per day, almost five times the reference value.

\textcolor{black}{Extrapolating the analysis from 10,000 meters to approximately 1.5 million meters results in a scaling factor of \(1{,}500{,}000/10{,}000 = 150\) \cite{noauthor_grid_nodate}. Under the same assumptions for daily energy consumption and communication degradation, the estimated economic exposure increases from approximately \$1,009 to \$151,300 per day at \$59.38/MWh and from \$5,097 to \$764,600 per day at \$300/MWh. These estimates illustrate a worst-case scenario rather than predict the actual utility losses, since utilities may use data estimation, communication recovery, and other mitigation measures under such critical conditions.} However, the results corroborate that cyberattacks targeting AMI communication infrastructure may have consequences beyond communication reliability, producing economically significant impacts on utility operations. 

\vspace{-0.2cm}
\section{Conclusions} \label{s:Conclusions}

In this work, the impact of communication degradation on AMI was studied under different cyberattack scenarios. A Mininet-based testbed was developed for this task and 
five scenarios were evaluated, i.e., communication delay, $20\%$ packet loss, $50\%$ packet loss, combined delay and packet loss, and data concentrator failure. 
The results demonstrate that communication delays increase latency, while packet loss significantly increases timeout rates and reduces measurement availability. The combined scenario produced the highest degradation, showing that delay and packet loss together have a stronger impact on communication performance. On the other hand, the data concentrator failure scenario demonstrated that the proposed dual redundancy architecture can maintain data availability. \looseness=-1

\textcolor{black}{In the economic analysis, the measured degradation rates were combined with ISO-NE electricity prices. For 10,000 affected meters, the estimated economic exposure was approximately \$1,009 per day at \$59.38/MWh and \$5,097 per day at \$300/MWh. When extrapolated to 1.5 million meters under the same assumptions, these estimates increased to approximately \$151,300 and \$764,600 per day, respectively. These values illustrate the impact that cyberattacks targeting AMI can potentially have in the worst-case scenario.} 

\vspace{-0.3cm}
\bibliographystyle{IEEEtran}
\bibliography{references}
\end{document}